\documentclass[a4paper,twocolumn,british,prl,nofootinbib]{revtex4-1}
\usepackage{lmodern}

\usepackage[T1]{fontenc}
\usepackage[latin9]{inputenc}
\usepackage{color}
\usepackage{babel}
\usepackage{booktabs}
\usepackage{amsmath}
\usepackage{amssymb}
\usepackage[unicode=true,pdfusetitle,
 bookmarks=true,bookmarksnumbered=false,bookmarksopen=false,
 breaklinks=false,pdfborder={0 0 0},backref=false,colorlinks=true]
 {hyperref}
\hypersetup{
 citecolor=blue,filecolor=blue,linkcolor=blue,urlcolor=blue}

\makeatletter

\pdfpageheight\paperheight
\pdfpagewidth\paperwidth

\providecommand{\tabularnewline}{\\}

\renewcommand{\[}{\begin{equation}}
\renewcommand{\]}{\end{equation}} 
\usepackage{array}

\makeatother

\begin{document}

\title{Limits of Quasi-Static Approximation in Modified-Gravity Cosmologies}

\author{Ignacy Sawicki}

\affiliation{Départment de physique théorique, Université de Genève, \\
Quai Ernest Ansermet 24, 1211 Genève, Switzerland}

\email{ignacy.sawicki@unige.ch}

\author{Emilio Bellini}

\affiliation{Institut de Ciènces del Cosmos, Universitat de Barcelona, IEEC-UB,
\\
Martì i Franquè 1, E08028 Barcelona, Spain}

\email{emilio.bellini@icc.ub.es}

\begin{abstract}
We investigate the limits of applicability of the quasi-static approximation
in cosmologies featuring general models of dark energy or modified
gravity. We show that, at best, the quasi-static approximation breaks
down outside of the \emph{sound} horizon of the dark-energy, rather
than the cosmological horizon as is frequently assumed. When the sound
speed of dark energy is significantly below that of light, the quasi-static
limit is only valid in a limited range of observable scales and this
must be taken into account when computing effects on observations
in such models. As an order of magnitude estimate, in the analysis
of data from today's weak-lensing and peculiar-velocity surveys, dark
energy can be modelled as quasi-static only if the sound speed is
larger than order 1\% of that of light. In upcoming surveys, such
as Euclid, it should only be used when the sound speed exceeds around
10\% of the speed of light. In the analysis of the cosmic microwave
background, the quasi-static limit should never be used for the integrated
Sachs-Wolf effect and for lensing only when the sound speed exceeds
10\% of the speed of light.
\end{abstract}
\maketitle

\section{Introduction}

Dynamical dark energy (DE) and modified gravity (MG), modelling the
observed acceleration of late-time cosmological expansion, usually
require a new degree of freedom beyond those present in standard $\Lambda$-cold
dark matter ($\Lambda$CDM) cosmology.\footnote{Although modifications of general-relativistic (GR) constraints also
exist, e.g.~refs~\cite{Afshordi:2006ad,Carroll:2006jn,Lim:2010yk}.} The full dynamics of this new degree of freedom in general models
of DE/MG are usually quite complicated. However, since galaxy-survey
data are still mostly available only on scales small compared to the
cosmological horizon, the quasi-static (QS) approximation is frequently
used to approximate the full DE/MG dynamics and interpret observations.
Roughly, this amounts to neglecting terms involving time derivatives
in the Einstein equations for perturbations and only keeping spatial
derivatives. Such a procedure is thought to be valid on sufficiently
small scales, typically assumed to be those well inside the cosmological
horizon. It is well known from numerical studies that in the case
of quintessence \cite{Ratra:1987rm,Wetterich:1987fm}, $f(R)$ gravity
\cite{Carroll:2003wy} or the covariant galileon models \cite{Nicolis:2008in,Deffayet:2009wt}
this approximation is good enough on observable linear scales (see
e.g.~\cite{Song:2006ej,DeFelice:2010as,Barreira:2012kk} and in particular
ref. \cite{Noller:2013wca}). On the other hand, general perfect-fluid
DE can show some discrepancy between the full and QS solutions at
late times \cite{Sapone:2009mb}. On non-linear scales, N-body simulations
have been used to study symmetron-screened \cite{Hinterbichler:2010es}
non-minimally coupled quintessence \cite{Llinares:2013qbh,Llinares:2013jua}
and $f(R)$ gravity \cite{Bose:2014zba} with largely the same conclusion:
the effects of the time-derivative terms are small. But how general
are these results?

The solutions in the dust-dark-energy dynamical system which models
the late universe can indeed always be described using two functions
of space and time \cite{Amendola:2007rr,Hu:2007pj,Bertschinger:2008zb,Zhao:2008bn,Pogosian:2010tj,Bean:2010zq,Sawicki:2012re,Baker:2012zs}
without reference to the extra degree of freedom. Taking the QS limit
allows one to remove the dependence on the initial conditions for
the DE/MG and to reinterpret what is a parameterisation of a particular
solution as a universal description of a model. In the extreme QS
limit, all the scale-dependence is also neglected and the DE model
is described using just two functions of time: the effective Newton's
constant $\mu$ and the gravitational slip parameter. Allowing for
scale dependence in principle gives a closer description on a wider
set of scales \cite{Amendola:2012ky,Silvestri:2013ne,Motta:2013cwa},
but the question arises to what extent all of these approximations
are valid.

Recently, effective-field-theory (EFT) methods \cite{Creminelli:2008wc,Gubitosi:2012hu,Bloomfield:2012ff,Piazza:2013pua,Gleyzes:2013ooa,Bellini:2014fua,Gleyzes:2014dya}
have been developed to efficiently encode the full dynamics of DE/MG
models featuring a single new degree of freedom. Linear structure
formation in a very general class of DE/MG models (Horndeski scalar-tensor
gravity \cite{Horndeski:1974wa,Deffayet:2011gz} and beyond \cite{Zumalacarregui:2013pma,Gleyzes:2014dya,Gao:2014soa})
can be fully described in this way using a small set of functions
of time only. We can thus use this formulation to compare QS and full
solutions in a general way in order to understand the approximations
properly. In what follows, we will use the term ``dark energy''
to refer to the models described by the EFT, i.e.~to also most models
of modified gravity.

In this paper, we show that the domain of validity of the QS approximation
is at best determined by the sound horizon of dark energy: even when
the parameters of the DE models evolve slowly, the QS approximation
can only be trusted inside the sound horizon. The scale of this sound
horizon can be calculated from an algebraic relationship between the
parameters of the EFT or, alternatively, is a functional of the action
for the DE model. We derive a condition determining when the QS approximation
can be used safely to analyse a particular dataset and apply this
result to parameters of current and future surveys.

\section{The Quasi-Static Solution\label{sec:QSsol}}

We model the late universe as a mixture of pressureless dust with
density fraction $\Omega_{\text{m}}$, representing both dark matter
and baryons between which we do not differentiate, and the dark energy
with density fraction $\Omega_{\text{DE}}=1-\Omega_{\text{m}}$, which
we assume comprises a single degree of freedom. The background expansion
history is constrained by observations (e.g. ref.~\cite{Planck:2015xua})
to be that corresponding to a source with approximately constant pressure
and no spatial curvature. In the linear perturbed Einstein equations
describing the evolution of structure on largest scales, an extra
scalar degree of freedom can be integrated out by solving the constraints.
This procedure, described in detail in ref.~\cite{Bellini:2014fua},
reduces the exact perturbation equations into a second-order differential
equation for the gravitational potential $\Phi$ coupled to the pressureless
matter. The full form for such an equation for general DE models which
do not contain operators violating Lorentz invariance is given in
\cite{Bellini:2014fua,Gleyzes:2014rba}.\footnote{The beyond-Horndeski models of ref.~\cite{Gleyzes:2014dya} contain
higher-derivative operators which cancel once the constraints are
solved and therefore their dynamics reduce to (\ref{eq:phieq}) in
the appropriate limit. Including true Lorentz-violating operators,
brings higher-order spatial derivatives, $k^{4}$-terms, to eq. (\ref{eq:phieq}).
These sort of terms would lead to a violation the $r^{-2}$ law in
gravity, which constrains them to lengthscales $l\lesssim1\,\mu\text{m}$
\cite{Blas:2015xga}. There must therefore be a large hierarchy between
the sound-speed and the $k^{4}$ terms. In such a case, including
such suppressed $k^{4}$ terms does not change any of the conclusions
of this paper.}\textbf{ }However, for the purpose of this work, it is enough to write
as a proxy an approximation valid at smaller scales, where it can
be written as
\begin{align}
\mathcal{E}\left[\Phi\right]\equiv\Phi'' & +\left(\gamma+\frac{H'}{H}\right)\Phi'+\left(\frac{M^{2}}{H^{2}}+c_{\text{s}}^{2}k_{H}^{2}\right)\Phi=\label{eq:phieq}\\
 & =-\frac{3c_{\text{s}}^{2}\mu_{\delta}}{2}\Omega_{\text{m}}\delta-\frac{3c_{\text{s}}^{2}\mu_{v}}{2}\Omega_{\text{m}}Hv\,,\nonumber 
\end{align}
where we use primes to denote derivatives w.r.t. e-folding time, $\delta$
and $v$ are the matter density contrast and velocity potential in
Newtonian gauge, $H$ is the Hubble parameter, $a$ the scale factor,
$k_{H}\equiv k/aH$ is the comoving mode number in units of the cosmological
horizon, while $\gamma$ is a model-dependent friction coefficient.
The sound speed $c_{\text{s}}>0$ is the speed at which high-frequency
longitudinal (scalar) perturbations in the DE propagate, while $M$
is some effective mass of the extra degree of freedom related to parameters
of the model and the background expansion history (see e.g.~\cite{Bellini:2014fua}).
The potential is coupled to matter through an effective Newton's constant,
$\mu_{\delta}$, and an equivalent coupling to the matter velocity
potential, $\mu_{v}$. We take both of these to be constant, but time
evolution does not add qualititative changes provided it be slow,
with timescale \textbf{$H$}. The system is completed by the standard
EMT conservation equations for dust
\begin{align}
\delta'=k_{H}^{2}Hv+3\Phi'\,,\qquad Hv'=-\Phi\,,\label{eq:cons}
\end{align}
where we have assumed that there is no gravitational slip, $\Phi=\Psi$.\footnote{In the EFT, this sets both the parameters $\alpha_{\text{T}}=\alpha_{\text{M}}=\alpha_{\text{H}}=0$
\cite{Saltas:2014dha}.} The two gravitational potentials are always related through a constraint
rather than a dynamical equation, and therefore generalising this
result to models with non-vanishing slip does not change qualitatively
the results presented below: if the quasi-static approximation is
good for $\Phi$, it is also good for $\Psi$. 

The coupling to the velocity $\mu_{v}$ always appears when the Einstein
$(00)$ and $(0i)$ equations are used to eliminate the DE scalar
perturbation and it does not in general disappear when the comoving-gauge
density perturbation is used. Secondly, we stress that since the modification
of gravity contains an additional scalar on top of the standard scalar
density perturbations in matter, the time derivatives of the potentials
$\Phi$ depend on two time-scales: (i) that of the pressureless collapse
of the dust, which is just $H$ and does not produce any oscillations
since there is no pressure (ii) of the modified-gravity scalar degree
of freedom, which is related to its mass and the sound horizon and
thus to the pressure support it provides. This is somewhat different
than in ref.~\cite{Baker:2014zva}, where the time-scale intrinsic
to the scalar was not considered to determine the oscillations of
the gravitational potentials.

In principle, to solve the system (\ref{eq:phieq}-\ref{eq:cons}),
we should find the eigenmodes and their initial conditions. This is
in general impossible. We instead can \emph{define }the QS solution
by splitting the Newtonian potential into an oscillating part $\varphi$
and the quasi-static part $\Phi_{\text{QS}}$
\begin{equation}
\Phi\equiv\varphi+\Phi_{\text{QS}}\,,\label{eq:QSdef}
\end{equation}
choosing $\Phi_{\text{QS}}$ such that the oscillating part $\varphi$
has no source in its equation of motion, $\mathcal{E}\left[\varphi\right]=0$,
with $\mathcal{E}$ denoting the homogeneous part of eq. (\ref{eq:phieq}).
This gives us one decoupled equation for $\varphi$ and the dependent
eqs. (\ref{eq:phieq}-\ref{eq:cons}). Since the equation for $\varphi$
is source free, provided that its coefficients are appropriate, the
oscillations $\varphi$ can be expected to decay away \emph{over time},
leaving us with only what we have called the QS solution. This is
close but not exactly the same definition as is used when time derivatives
are just neglected. We include extra information which allows us to
make a definite statement about the region of validity, inside of
which the two solutions are the same. 

As we will see, there are some corrections to $\mathcal{E}\left[\varphi\right]$
which are suppressed by $(c_{\text{s}}k_{H})^{-2}$. These corrections
will therefore place an ultimate limit to the scales on which the
QS limit can be considered to be valid.

\section{Maximum Domain of Validity\label{sec:domain}}

In general, we cannot find an exact solution for the particular integral
of (\ref{eq:phieq}). We can, however, obtain the Padé approximant
to this solution, valid at least on some scales. Therefore, for the
QS solution, we will be seeking an approximate solution valid inside
some set of small scales to some accuracy.

We start off by considering the typical evolution of cosmological-perturbation
variables inside the cosmological horizon, $k_{H}\gg1$. In the usual
case, we have $Hv\sim\mathcal{O}(k_{H}^{-2}\delta)$ by virtue of
the continuity equation (\ref{eq:cons}). This gives a natural hierarchy
for these variables which will allow us to carry out an expansion.
In what follows, we will therefore consider
\[
\delta\sim\mathcal{O}(\epsilon^{0})\,,\qquad Hv\sim\mathcal{O}(\epsilon)\,,
\]
identifying the order parameter $\epsilon$ roughly with $k_{H}^{-2}$.
This also gives us $\Phi\sim\mathcal{O}(\epsilon)$. By construction,
our approximation always fails at the cosmological horizon, if not
before.

Given this hierarchy, we can choose an ansatz for the QS solution,
which includes the contributions lowest-order in $\epsilon$ and is
compatible with eq.~(\ref{eq:phieq}), 
\begin{equation}
k_{H}^{2}\Phi_{\text{QS}}\equiv-\frac{3}{2}\left(A_{1}+\frac{\epsilon A_{2}}{k_{H}^{2}}\right)\Omega_{\text{m}}\delta-\frac{3\epsilon B_{1}}{2}\Omega_{\text{m}}Hv\,.\label{eq:PhiQS}
\end{equation}
The time-dependent coefficients $A_{1},A_{2,}B_{1}$ are chosen in
such a way that eq.~(\ref{eq:phieq}) can be written as $\mathcal{E}\left[\varphi\right]=0$
up to the relevant precision defined by $\epsilon$ when $\Phi$ is
replaced according to eq.~(\ref{eq:QSdef}). 

In the extreme quasi-static limit, $k_{H}\rightarrow\infty$, one
typically neglects the $A_{2}$ and $B_{1}$ terms, since they are
suppressed by $\epsilon$. However, if we wish to include, for example,
the corrections resulting from the mass $M^{2}$, one needs to include
terms such as $A_{2}$ and therefore also $B_{1}$ since both of these
corrections appear at the same order. Moreover, as we will see, including
$B_{1}$ allows us to estimate the domain of validity for the QS solution.\\

One now substitutes the QS ansatz (\ref{eq:QSdef}) into eqs (\ref{eq:phieq}),
(\ref{eq:cons}) and (\ref{eq:QSdef}), keeping all terms involving
$\varphi$ and replacing any time derivatives of $\delta$ and $v$
using the conservation equations. The evolution equation (\ref{eq:phieq})
can then be rewritten as 
\begin{align}
\varphi''+ & \left(\gamma+\frac{H'}{H}\right)\varphi'+\label{eq:varphievol}\\
 & +\left(\frac{M^{2}}{H^{2}}+\frac{3}{2}A_{1}(1+3c_{\text{s}}^{2})\Omega_{\text{m}}+\frac{c_{\text{s}}^{2}k_{H}^{2}}{\epsilon}\right)\varphi=\nonumber \\
= & \left(\frac{\mathcal{F}_{1}}{\epsilon}+\frac{\mathcal{F}_{2}}{k_{H}^{2}}\right)\Omega_{\text{m}}\delta+\frac{3}{2}\mathcal{F}_{3}\Omega_{\text{m}}Hv+\mathcal{O}(\epsilon)\,,\nonumber 
\end{align}
with the $\mathcal{F}_{i}$ functions of time. According to how we
have defined the QS solution, we need to pick such coefficients $A_{1,2}$
and $B_{1}$ that all the $\mathcal{F}_{i}=0$. The particular choices
we require are:
\begin{eqnarray}
A_{1} & = & \mu_{\delta}\,,\label{eq:QSsol}\\
c_{\text{s}}^{2}A_{2} & = & -\frac{3}{2}\Omega_{\text{m}}\mu_{\delta}^{2}-\mu_{\delta}\left(1-\frac{H'}{H}+\frac{M^{2}}{H^{2}}-\gamma\right)\,,\nonumber \\
c_{\text{s}}^{2}B_{1} & = & c_{\text{s}}^{2}\mu_{v}+\mu_{\delta}(4-\gamma)\,.\nonumber 
\end{eqnarray}
This choice fixes all the freedom in ansatz (\ref{eq:PhiQS}) and
removes the external source in the evolution equation for $\varphi$,
eq.~(\ref{eq:varphievol}), decoupling it from the matter perturbations.
This choice also matches the standard choice for the leading-order
terms of the QS solution. In this new description, $\varphi$ evolves
with no source and, given certain additional conditions, $\varphi$
decays away and the QS solution is eventually reached. We discuss
these additional conditions in section~\ref{sec:oscillation}. \\

Let us now turn to the evolution equation for $\delta$, or equivalently,
the equation for the growth factor, which is usually the equation
of interest when the QS approximation is used. Using the QS ansatz
given by (\ref{eq:PhiQS}) with the coefficient values given in eqs
(\ref{eq:QSsol}), we eliminate the matter velocity potential $v$
from the matter conservation equations (\ref{eq:cons}), obtaining
\begin{equation}
\delta''+\left(2+\frac{H'}{H}+\epsilon\mathcal{F}_{4}\right)\delta'-\frac{3}{2}\mu_{\delta}\Omega_{\text{m}}\left(1+\epsilon\mathcal{F}_{5}\right)\delta=S[\varphi]\label{eq:growthfactor}
\end{equation}
where the source is determined by the deviation away from the QS solution,
\begin{align*}
S[\varphi] & \equiv-\left(k_{H}^{2}-\frac{9}{2}\epsilon\mu_{\delta}\Omega_{m}\right)\varphi+3\epsilon\left(2+\frac{H'}{H}\right)\varphi'+3\epsilon\varphi'',
\end{align*}
and
\begin{align*}
c_{\text{s}}^{2}k_{H}^{2}\mathcal{F}_{4} & \equiv\Omega_{\text{m}}\left(\frac{3}{2}\mu_{\delta}(\gamma-4)\right)+\mathcal{O}(c_{\text{s}}^{2})\,,\\
c_{\text{s}}^{2}k_{H}^{2}\mathcal{F}_{5} & \equiv\frac{M^{2}}{H^{2}}+\frac{3}{2}\mu_{\delta}\Omega_{\text{m}}+1-\frac{H'}{H}-\gamma+\mathcal{O}(c_{\text{s}}^{2})\,.
\end{align*}
In the small scale limit of eq. (\ref{eq:growthfactor}), $k_{H}\rightarrow\infty$,
we recover the standard equation for the growth factor with sources
which all involve the oscillating part of the potential, $\varphi$.
Thus \emph{if }indeed we can show that $\varphi$ decays away at these
scales, the QS solution for the growth factor will be valid once that
has occurred. One should note that as the mode evolves from superhorizon
to subhorizon the deviation from QS $\varphi$ does contribute to
the dynamics. Thus even if deep inside the sound horizon the QS relationship
between $\Phi$ and $\delta$ is correct, it is not necessarily true
that the standard relationship between the superhorizon curvature
perturbation and the subhorizon $\delta$ is maintained (see section
\ref{sec:oscillation} for a detailed discussion).

As we move away from the extremely small scales, corrections to the
evolution appear. In particular, if $M^{2}\gg H^{2}$, then the corrections
due to this mass are as expected from the form of the coefficient
of $\Phi$ in eq.~(\ref{eq:phieq}) and as is currently taken in
$f(R)$ gravity models \cite{Tsujikawa:2007xu}. There are also some
additional corrections proportional to $\mu_{\delta}H^{2}$ which
can only become relevant when the QS approximation in breaking down
close to the cosmological horizon. Thus the naive QS estimate of the
effective Newton's constant is essentially valid at cosmological subhorizon
scales.

Crucially, there are also corrections to the friction term, $\delta'$.
The corrections are of order $\mu_{\delta}$ and are only suppressed
by the sound horizon for the dark energy, $c_{\text{s}}k_{H}$ rather
than the cosmological horizon $k_{H}$. Thus, at scales near the\emph{
sound horizon,} the response of the growth function to some particular
value of the effective Newton's constant is different than what would
be expected from the naive guess, \emph{even }if the quasi-static
solution for $\Phi$ were valid. The typical sign of this correction
is negative, thus the growth rate increases as the sound horizon is
approached. This suppression by only the sound horizon is driven by
the fact that the solution for the coefficient $B_{1}$ contains terms
of order $c_{\text{s}}^{-1}$ and thus such corrections are missed
when the dust velocity potential $v$ is ignored.\\

We should also add that similar corrections appear to the evolution
equation for $\varphi$, eq.~(\ref{eq:varphievol}) at order $\epsilon$.
The coefficient of the friction term is also modified by a term $\mathcal{F}_{4}/c_{\text{s}}^{2}k_{H}^{2}$,
thus leading to a similar decrease in the friction and therefore a
relative enhancement of the amplitude of the oscillation $\varphi$.
Thus the sound horizon also provides a limiting scale for the quasi-static
approximation to hold for the potential $\Phi$.

We also remind the reader that in the derivation above we have neglected
all time derivatives of the DE model parameters. If the time scale
of the evolution is of order $H$, then including these terms does
not qualitatively affect the results. However, if the model features
a rapid time evolution of its parameters, e.g.~with timescale $H/\epsilon$,
then these corrections can be large and would dramatically change
the conclusions. Unsurprisingly, if the model features a rapid time
evolution of its properties, one cannot hope that a quasi-static approximation
can hold.

\section{Evolution of Oscillations\label{sec:oscillation}}

In the previous section, we demonstrated that, inside the dark-energy
sound horizon, the quasi-static solution can in principle model the
evolution of the full dark-energy-dust system. Choosing a scale well
inside the sound horizon, $c_{\text{s}}k_{H}\gg1$, is a necessary
condition for the existence of this QS solution. It is not sufficient,
however. In addition, the deviation from the QS solution $\varphi$
must decay faster than the QS solution, such that the QS solution
is asymptotically reached after sound-horizon crossing. In this section,
we will demonstrate the conditions required for this decay to occur.

The definition of the QS solution we have proposed in section \ref{sec:QSsol}
was constructed in such a way that the variable $\varphi$ --- describing
the deviation of the gravitational potential from the QS solution
--- has no source in its evolution equation up to some precision.
This decouples the evolution of $\varphi$ from the dust part of the
system and therefore its behaviour can be considered in isolation.
Taking the small-scale limit, $k_{H}\rightarrow\infty$, where the
QS solution should be valid, if it is valid anywhere at all, the variable
$\varphi$ obeys the homogeneous part of eq.~(\ref{eq:varphievol}), 

\begin{equation}
\ddot{\varphi}+(\gamma-1)\mathcal{H}\dot{\varphi}+c_{\text{s}}^{2}k^{2}\varphi=0\,,\label{eq:varphievol-1}
\end{equation}
where we have re-expressed eq.~(\ref{eq:varphievol}) in conformal
time $\eta$, with the overdot signifying a derivative w.r.t.~it,
while $\mathcal{H}\equiv aH$ is the conformal Hubble parameter. 

For any particular model with its appropriate background, eq. (\ref{eq:varphievol-1})
can be solved numerically. Here, we will only perform an estimate
in a situation where closed-form solutions are available: assuming
power-law evolution where necessary. The conformal Hubble parameter
evolves as $\mathcal{H}=2/\left|1+3w_{\text{tot}}\right|\eta$ where
$w_{\text{tot}}\neq-1/3$ is the total equation of state for the universe.
We are also going to assume that the dark-energy sound speed evolves
as a power law with redshift, $c_{\text{s}}^{2}=c_{s0}^{2}(1+z)^{-p}=c_{s0}^{2}\eta^{2p/\left|1+3w_{\text{tot}}\right|}\,.$
We note here that for $p<-1$, the comoving sound horizon would actually
be shrinking during matter domination and therefore the modes would
be leaving the sound horizon. For any particular mode, the super-sound-horizon
corrections to eq. (\ref{eq:varphievol-1}) would eventually take
over and the QS limit would be violated completely by the results
of section~\ref{sec:domain}.

The friction term $\gamma$ for general scalar-tensor models is determined
by the action and given in full in ref.~\cite{Bellini:2014fua}.
In MG models, where the kineticity is small compared to the braiding,
$\alpha_{\text{K}}\ll\alpha_{\text{B}}$ (e.g. $f(R)$),\textbf{ $\gamma\approx3$};
on the other hand, in perfect-fluid DE models, such as k-essence/quintessence
$\gamma\approx4$. This value is model-dependent, so we will only
assume it is a constant during the era being considered. We can then
rewrite eq.~(\ref{eq:varphievol-1}) in (one of the) standard forms
of the Bessel equation, the solutions of which are oscillatory with
a decaying envelope. Expressing the result as a function of redshift,
the envelope evolves as $\varphi_{\text{env}}\propto(1+z)^{\nu}$
with decay exponent
\begin{equation}
\nu=\frac{p}{4}+\frac{\gamma-1}{2}\,,\label{eq:decayexpo}
\end{equation}
when inside the sound horizon. Note that this is independent of $w_{\text{tot}}$
and is also a good approximation for the $\Lambda$CDM background.
So if we take $\gamma=3$, then $p>-4$ for a decaying $\varphi$.
Thus the sound speed must decrease rapidly for $\varphi$ to grow.
Beware that the comoving cosmological horizon shrinks during acceleration
($z<0.65$ in $\Lambda$CDM with $\Omega_{\text{m}}=0.31$) and the
modes begin to exit the sound horizon, potentially invalidating the
QS limit in any case. 

In eq.~(\ref{eq:varphievol-1}) we have neglected the mass term $M^{2}a^{2}$.
For example in $f(R)$ models, this mass can be larger than the sound-speed
term even on scales inside the sound horizon. On these scales, the
evolution of $\varphi_{\text{env}}$ is determined by the evolution
of the mass, but the solution goes through similarly provided it is
a power law in $(1+z)$. The exponent $\nu$ will change as the mode
crosses the scale $k(1+z)\sim M/c_{\text{s}}$.\\

The conditions we have derived refer to the QS relationship between
the gravitational potential and the matter density perturbation, as
given by eq. (\ref{eq:PhiQS}). A separate issue is the relationship
between the initial superhorizon curvature perturbation, as set up
by inflation, and the amplitude of the mode inside the horizon. \emph{Every
}mode starts off superhorizon and will have evolved through scales
where the QS limit fails, even if the scales probed by a the survey
are in the deep QS regime. The validity of the QS solution guarantees
the evolution for the growth factor from one redshift to another,
but does not necessarily ensure that the amplitude will not have been
modified previously by the interaction between the DE degree of freedom
and the matter on superhorizon scales and while the deviation from
QS is dying off. Eq.~(\ref{eq:growthfactor}) contains a source involving
$\varphi$ for a large fraction of the evolution history of all modes.
This will end up having an impact on the shape and normalisation of
the matter power spectrum inside the sound horizon which in a generically
scale-dependent manner. See ref. \cite{Kunz:2015oqa} for an explicit
example. The only way to suppress this effect is to ensure that for
the modes being observed, the modifications of gravity were insignificant
at the time that the modes were outside of their sound horizon. 

We discuss the impact of the above results on current and upcoming
surveys in the next section.

\section{Application to Surveys\label{sec:surveys}}

In the preceding, we have built up a picture where, outside the dark-energy
sound horizon, the configuration of the dark energy degree of freedom
is significantly different from the QS solution. Provided that the
comoving sound horizon be growing, the mode will cross it. The deviation
from the QS solution will then decay away with the exponent given
by eq. (\ref{eq:decayexpo}), \emph{eventually }becoming negligible
given some required theoretical precision. We will ignore the possible
temporary effect of this decaying deviation on the density perturbation
in this section.

Given the above, we can estimate the minimum dark-energy sound speed
which is necessary for the QS approximation to be appropriate in the
analysis of observations given particular survey parameters. We will
make the assumption that the QS limit of the dark-energy model is
very simple, with no rapid time evolution of its properties or scale
dependence. This is typical of the models which are considered in
data analysis (e.g. ref. \cite{Samushia:2013yga}). We assume that
\begin{itemize}
\item the cosmological background is exactly $\Lambda$CDM with parameters
approximating the Planck 2015 results, $\Omega_{\text{m0}}=0.31$
and $z_{\text{eq}}=3371$ \cite{Planck:2015xua};
\item the sound speed of dark energy $c_{\text{s}}$ is constant, $p=0$;
we take a model with $\gamma=3$, i.e.~with the exponent of the decay
of oscillations of eq.~(\ref{eq:decayexpo}) constant, $\nu=1$;
\item the value of the effective Newton's constant in the QS solution is
in principle redshift dependent and given by $\mu_{\delta}(z)$;\footnote{Allowing for redshift dependence of $\mu_{\delta}$ does not change
$A_{1}$ in eq.~(\ref{eq:QSsol}), which is all that is necessary
for the discussion here.}
\item the survey uses some longest mode $k_{\text{min}}\equiv2\pi/r_{\text{max}}$
for their DE-related analysis using a sample of galaxies within a
redshift bin centred on $z_{\text{surv}}$.
\end{itemize}
We then require that, for the longest mode $k_{\text{min}}$, the
oscillating part of the potential constitutes no more than fraction
$\varepsilon$ of the QS value, i.e.~the DE sound speed needs to
be large enough to satisfy 
\begin{equation}
\frac{\varphi(z_{*})}{\Phi_{\text{\text{QS}}}(z_{\text{surv}})}\left[\frac{1+z_{\text{surv}}}{1+z_{*}}\right]^{\nu}\lesssim\varepsilon\label{eq:inQS}
\end{equation}
where $z_{*}$ is the redshift at which the mode $k_{\text{min}}$
crossed the DE sound horizon, obtained by solving $c_{s}k_{\text{min}}=\mathcal{H}(z_{*})$.
The deviation from the QS solution at sound-horizon crossing can be
roughly estimated by assuming that above its sound horizon, the dark
energy will behave roughly as dust (since there is no pressure support),
so one would expect $k_{H}^{2}\Phi\sim-\frac{3}{2}(\Omega_{\text{m}}+\Omega_{\text{DE}})\delta$.
On the other hand, the QS result is given by $k_{H}^{2}\Phi_{\text{QS}}=-\frac{3}{2}\mu_{\delta}\Omega_{\text{m}}\delta$.
Thus 
\[
\varphi(z_{*})=\Phi-\Phi_{\text{QS}}\sim-\mu_{\delta}^{-1}\left(\mu_{\delta}-1-\frac{\Omega_{\text{DE}}}{\Omega_{\text{m}}}\right)\Phi_{\text{QS}}(z_{*})\,.
\]
Let us now parameterize the effective Newton's constant as \cite{Bertschinger:2008zb}
\[
\mu_{\delta}=1+\mu_{0}(1+z)^{-s}
\]
and assume that $\mu_{\delta}$ is not too different from one, so
that the potential is approximately constant during matter domination.
Also, we assume that the dark energy is not of tracking type and therefore
$\Omega_{\text{DE}}(z_{*})\ll\Omega_{\text{m}}(z_{*})$. In order
to obtain an analytic solution, we will also approximate the Hubble
parameter as $\mathcal{H}\approx H_{0}\sqrt{\Omega_{\text{m0}}(1+z)}$,
which is a reasonable approximation during matter domination. Putting
the above together, we obtain from condition (\ref{eq:inQS})
\begin{equation}
c_{\text{s}}\gtrsim\sqrt{\Omega_{\text{m0}}}\frac{H_{0}}{k_{\text{min}}}\left(\frac{\mu_{0}}{\epsilon}\right)^{1/(s+\nu)}(1+z_{\text{surv}})^{\nu/(s+\nu)}\,.\label{eq:cslim}
\end{equation}
This should be thought of as an order-of-magnitude approximation,
which should be compared with numerical results in table~\ref{tab:reqsoncs}.
We can easily see that if the QS solution is to be good enough, we
need higher sound speeds for high-redshift surveys, large maximum
lengthscales surveyed, larger deviations from $\mu_{0}=0$ and more
slowly evolving effective Newton's constants, just as one would naively
expect. The minimum sound speed is weakly dependent on the survey
redshift, but strongly on the lowest mode $k{}_{\text{min}}$ being
used in the analysis. Eq.~(\ref{eq:cslim}) is the main result of
this paper, allowing for a simple estimate as to whether the QS approximation
can be used for a given dataset and DE model. We remind the reader
that this approximation is valid for simple DE models with constant
sound speed and a monotonic evolution of the effective Newton's constant.

In table~\ref{tab:reqsoncs}, we have summarised a selection of surveys
and the constraints on the sound speed given the requirement of quasi-staticity.
It shows that today's galaxy surveys are well inside the region where
the QS limit is valid for our simple models with sound speed $c_{\text{s}}>0.02$.
There is an interesting effect whereby, for the surveys centred on
lower redshifts, such as BOSS, the QS approximation is better than
expected since the comoving horizon shrinks for $z<0.65$ and therefore
the modes $k_{\text{min}}$ entered the sound horizon at higher redshift
and are already leaving it at the time of the survey. Once Euclid
data are available, the quasi-static approximation will only be appropriate
for dark-energy models with sound speeds close to that of light. 

On the other hand, the CMB anisotropies already probe very large scales,
since the Hubble horizon at recombination is approximately one degree
on the sky today. This means that if the DE was playing any role at
all at recombination (e.g. in Early Dark Energy models \cite{Doran:2006kp,Pettorino:2013ia})
the quasi-static limit cannot be consistently used there, even for
models with $c_{\text{s}}=1$. Whether it is consistent to use the
QS approximation for late time DE in CMB data is, however, subtle.
One must ask both which range of scales $k$ contribute to the kernel
of $C_{\ell}$ for the particular multipole of interest $\ell$ and
whether at the redshift at which they contribute, the DE perturbations
are at all important. Only if they are, then one must consider whether
the perturbations of the DE and whether they are already quasistatic
at those scales. 

The DE perturbations have two main effects on the CMB: a modification
of the integrated Sachs-Wolf effect (ISW) and the lensing of the CMB.
the full the ISW effect has its kernel peaked at approximately $k=0.004$~Mpc$^{-1}$
for $\ell=20$ and $k=0.0005$~Mpc$^{-1}$ for $\ell=2$ and it is
sensitive to exactly the late times. Thus one should not use the QS
solution at all (see e.g.~\cite[fig. 4]{Song:2006jk}). Nonetheless,
the impact on parameter estimation is limited because of cosmic variance
at such scales.

On the other hand, Planck is sensitive to CMB lensing in multipoles
$40<\ell<400$ \cite{Ade:2015zua}. Using $\Lambda$CDM results as
a proxy, for $\ell=40$, half of the signal is from $z<1$; even at
$\ell=400$, half is still from $z<2$ \cite{Lewis:2006fu}. These
are the times at which DE dominates and therefore here also one must
be certain that the DE perturbations are calculated properly at the
scales of concern. For $\ell=40$, the kernel is peaked at $k=0.01$~Mpc$^{-1}$,
while for $\ell=400$, the peak is somewhat below $k=0.1$~Mpc$^{-1}$.
This implies that one should not use the QS approximation in CMB lensing
predictions whenever $c_{\text{s}}\lesssim0.1$. 

\begin{table}
\centering{}%
\begin{tabular}{lrrrr}
\toprule 
\multicolumn{5}{l}{\textsf{\textbf{Minimal sound speed for good QS limit}}}\tabularnewline
\textsf{\textbf{\footnotesize{}Survey:}} & \textsf{\textbf{\footnotesize{}BOSS}} & \textsf{\textbf{\footnotesize{}CFHTLenS}} & \textsf{\textbf{\footnotesize{}DES}} & \textsf{\textbf{\footnotesize{}Euclid}}\tabularnewline
\midrule
\midrule 
$z_{\text{surv}}$ & 0.57 & 0.75 & 0.6 & 2\tabularnewline
\midrule 
$k_{\text{min}}/h\text{ Mpc}^{-1}$ & 0.04 & 0.2 & 0.01 & 0.007\tabularnewline
\midrule 
$k_{\text{min}}/\mathcal{H}(z_{\text{surv}})$ & 140 & 700 & 50 & 20\tabularnewline
\midrule 
$c_{\text{s},\text{min}}@(\varepsilon/\mu_{0}=0.01,s=1)$ & 0.02 & 0.003 & 0.04 & 0.1\tabularnewline
\midrule 
$c_{\text{s},\text{min}}@(\varepsilon/\mu_{0}=0.01,s=3)$ & 0.009 & 0.002 & 0.02 & 0.06\tabularnewline
\bottomrule
\end{tabular}\caption{Minimal dark energy sound speed $c_{\text{s}}$ which allows for the
use of the quasi-static approximation for given example survey parameters
assuming $c_{\text{s}}$ be constant. Current surveys can be safely
analysed using the QS solution for $c_{\text{s}}\gtrsim0.02$ with
accuracy better than $\epsilon/\mu_{0}=1\%$. Future surveys will
probe much closer to the cosmological horizon and this sort of accuracy
can only be achieved for $c_{\text{s}}\gtrsim0.1$, depending on the
largest modes included in the analysis. When modes comparable to the
horizon are included, the QS limit should not be used at all. \protect \\
BOSS assumes largest scales used in redshift-space-distortion measurements
of $r_{\text{max}}=152h^{-1}$~Mpc \cite{Samushia:2013yga}. For
CFHTLenS, we assume largest angular scales available of $\theta_{\text{max}}=100'$
\cite{Simpson:2012ra}, while for DES --- that scales 3 times smaller
than the maximum observed in the 5000 sq. degree survey will be used,
i.e.~$23\protect\textdegree$ \cite{Abbott:2005bi}. Maximum Euclid
scale with significant data chosen as in ref.~\cite{Amendola:2013qna},
i.e.~$40\protect\textdegree$, which is approximately 1/3 of the
maximum angular scale available. Actual precision of QS for Euclid
is worse since cosmological horizon corrections are already significant
at those scales.We employ a full numerical computation rather than
the approximate result (\ref{eq:cslim}). \label{tab:reqsoncs} }
\end{table}

\section{Discussion and Conclusions}

Adopting the QS approximation essentially means that the DE degree
of freedom with a finite propagation speed is replaced by a modification
of the constraints of general relativity, i.e.~with instantaneous
propagation. The advantage is that the dynamics are much simpler,
insensitive to DE initial conditions and expressible in terms of simple
parameters. However, this approximation must fail whenever causality
places a limit, i.e.~beyond the dark energy sound horizon. We have
shown that for current galaxy surveys, the QS limit is sufficient
in the analysis whenever the DE model has sound speed $c_{\text{s}}\gtrsim0.02$.
As we enter the era of much wider surveys, such as Euclid, this approximation
becomes inappropriate even for models with sound speeds close to those
of light and could result in misleading constraints, both because
the sound horizon is too close to the scales involved in the analysis
and because the closeness of the cosmological horizon invalidates
the QS approximation in any case. Moreover, the CMB probes the universe
at the largest scales already today. Our result implies that the QS
limit is never valid for early dark energy, good enough for CMB lensing
if $c_{\text{s}}>0.1$ and always inappropriate for ISW since scales
close to the cosmological horizon are relevant.

On the other hand, most of the well-studied models of dark energy,
e.g~quintessence and $f(R)$, have sound speed equal to that of light.
The QS predictions for ISW in these models are in all likelihood somewhat
off the full solution, but we readily admit that cosmic variance at
those scales means that the biasing of parameter estimation is small.
For other probes, using the QS approximation in these particular models
is good enough for current surveys, as shown in the literature previously.
Thus the effects such as a scale-dependent Newton's constant in $f(R)$
gravity when $M^{2}\gg k^{2}/a^{2}$ do exist and might possibly be
observed \cite{Tsujikawa:2007xu}.

We have provided condition (\ref{eq:cslim}) as a simple test allowing
us to determine whether the use of the QS approximation could at all
be valid when we use observational data to test models of gravity
beyond the simplest ones such as $f(R)$ and quintessence --- for
example, those defined by their EFT parameters. This test should be
treated as a best-case scenario, which would be violated by a rapid
evolution of any of the parameters. One should also bear in mind that
there may appear a scale-dependent modification to the amplitude of
the matter power spectrum resulting from some dynamics during DE sound-horizon
crossing, even for modes which are well inside their QS regime during
observations (see ref.~\cite{Kunz:2015oqa}). As we map larger scales
in the near future with Euclid, LSST and the SKA, an EFT-like formulation
including all the time-domain behaviour will increasingly become necessary
to connect consistently initial conditions at large scales (e.g.~as
given by the the cosmic microwave background) with late-time observables.
Fully implementing the dynamics need not necessarily lead to a degradation
of constraints, but has been demonstrated to sometimes improve them
\cite{Hu:2013twa}.

\section*{Acknowledgements}

The work of E.B. is supported by the European Research Council under
the European Community\textquoteright s Seventh Framework Programme
FP7- IDEAS-Phys.LSS 240117. I.S. is supported by the Marie Sk\l odowska-Curie
Intra-European Fellowship Project ``DRKFRCS''. The authors are grateful
to Luca Amendola, Mustafa A. Amin, Martin Kunz, Mariele Motta, Federico
Piazza, Ippocratis Saltas, Licia Verde and Filippo Vernizzi for valuable
comments and criticism. 

\bibliographystyle{utcaps}
\bibliography{QSRefs}

\providecommand{\href}[2]{#2}\begingroup\raggedright\begin{thebibliography}{10}

\bibitem{Afshordi:2006ad}
N.~Afshordi, D.~J. Chung, and G.~Geshnizjani, ``{Cuscuton: A Causal Field
  Theory with an Infinite Speed of Sound},''
  \href{http://dx.doi.org/10.1103/PhysRevD.75.083513}{{\em Phys.Rev.}
  {\bfseries D75} (2007) 083513},
\href{http://arxiv.org/abs/hep-th/0609150}{{\ttfamily arXiv:hep-th/0609150
  [hep-th]}}.

\bibitem{Carroll:2006jn}
S.~M. Carroll, I.~Sawicki, A.~Silvestri, and M.~Trodden, ``{Modified-Source
  Gravity and Cosmological Structure Formation},''
  \href{http://dx.doi.org/10.1088/1367-2630/8/12/323}{{\em New J.Phys.}
  {\bfseries 8} (2006) 323},
\href{http://arxiv.org/abs/astro-ph/0607458}{{\ttfamily arXiv:astro-ph/0607458
  [astro-ph]}}.

\bibitem{Lim:2010yk}
E.~A. Lim, I.~Sawicki, and A.~Vikman, ``{Dust of Dark Energy},''
  \href{http://dx.doi.org/10.1088/1475-7516/2010/05/012}{{\em JCAP} {\bfseries
  1005} (2010) 012}, \href{http://arxiv.org/abs/1003.5751}{{\ttfamily
  arXiv:1003.5751 [astro-ph.CO]}}.

\bibitem{Ratra:1987rm}
B.~Ratra and P.~Peebles, ``{Cosmological Consequences of a Rolling Homogeneous
  Scalar Field},''
\href{http://dx.doi.org/10.1103/PhysRevD.37.3406}{{\em Phys.Rev.} {\bfseries
  D37} (1988) 3406}.

\bibitem{Wetterich:1987fm}
C.~Wetterich, ``{Cosmology and the Fate of Dilatation Symmetry},''
\href{http://dx.doi.org/10.1016/0550-3213(88)90193-9}{{\em Nucl.Phys.}
  {\bfseries B302} (1988) 668}.

\bibitem{Carroll:2003wy}
S.~M. Carroll, V.~Duvvuri, M.~Trodden, and M.~S. Turner, ``{Is cosmic speed -
  up due to new gravitational physics?},''
  \href{http://dx.doi.org/10.1103/PhysRevD.70.043528}{{\em Phys.Rev.}
  {\bfseries D70} (2004) 043528},
\href{http://arxiv.org/abs/astro-ph/0306438}{{\ttfamily arXiv:astro-ph/0306438
  [astro-ph]}}.

\bibitem{Nicolis:2008in}
A.~Nicolis, R.~Rattazzi, and E.~Trincherini, ``{The galileon as a local
  modification of gravity},''
  \href{http://dx.doi.org/10.1103/PhysRevD.79.064036}{{\em Phys. Rev.}
  {\bfseries D79} (2009) 064036},
\href{http://arxiv.org/abs/0811.2197}{{\ttfamily arXiv:0811.2197 [hep-th]}}.

\bibitem{Deffayet:2009wt}
C.~Deffayet, G.~Esposito-Farese, and A.~Vikman, ``{Covariant Galileon},''
  \href{http://dx.doi.org/10.1103/PhysRevD.79.084003}{{\em Phys. Rev.}
  {\bfseries D79} (2009) 084003},
\href{http://arxiv.org/abs/0901.1314}{{\ttfamily arXiv:0901.1314 [hep-th]}}.

\bibitem{Song:2006ej}
Y.-S. Song, W.~Hu, and I.~Sawicki, ``{The Large Scale Structure of f(R)
  Gravity},'' \href{http://dx.doi.org/10.1103/PhysRevD.75.044004}{{\em
  Phys.Rev.} {\bfseries D75} (2007) 044004},
\href{http://arxiv.org/abs/astro-ph/0610532}{{\ttfamily arXiv:astro-ph/0610532
  [astro-ph]}}.

\bibitem{DeFelice:2010as}
A.~De~Felice, R.~Kase, and S.~Tsujikawa, ``{Matter perturbations in Galileon
  cosmology},'' \href{http://dx.doi.org/10.1103/PhysRevD.83.043515}{{\em
  Phys.Rev.} {\bfseries D83} (2011) 043515},
  \href{http://arxiv.org/abs/1011.6132}{{\ttfamily arXiv:1011.6132
  [astro-ph.CO]}}.

\bibitem{Barreira:2012kk}
A.~Barreira, B.~Li, C.~M. Baugh, and S.~Pascoli, ``{Linear perturbations in
  Galileon gravity models},''
  \href{http://dx.doi.org/10.1103/PhysRevD.86.124016}{{\em Phys.Rev.}
  {\bfseries D86} (2012) 124016},
\href{http://arxiv.org/abs/1208.0600}{{\ttfamily arXiv:1208.0600
  [astro-ph.CO]}}.

\bibitem{Noller:2013wca}
J.~Noller, F.~von Braun-Bates, and P.~G. Ferreira, ``{Relativistic scalar
  fields and the quasistatic approximation in theories of modified gravity},''
  \href{http://dx.doi.org/10.1103/PhysRevD.89.023521}{{\em Phys.Rev.}
  {\bfseries D89} no.~2, (2014) 023521},
\href{http://arxiv.org/abs/1310.3266}{{\ttfamily arXiv:1310.3266
  [astro-ph.CO]}}.

\bibitem{Sapone:2009mb}
D.~Sapone and M.~Kunz, ``{Fingerprinting Dark Energy},''
  \href{http://dx.doi.org/10.1103/PhysRevD.80.083519}{{\em Phys.Rev.}
  {\bfseries D80} (2009) 083519},
\href{http://arxiv.org/abs/0909.0007}{{\ttfamily arXiv:0909.0007
  [astro-ph.CO]}}.

\bibitem{Hinterbichler:2010es}
K.~Hinterbichler and J.~Khoury, ``{Symmetron Fields: Screening Long-Range
  Forces Through Local Symmetry Restoration},''
  \href{http://dx.doi.org/10.1103/PhysRevLett.104.231301}{{\em Phys.Rev.Lett.}
  {\bfseries 104} (2010) 231301},
\href{http://arxiv.org/abs/1001.4525}{{\ttfamily arXiv:1001.4525 [hep-th]}}.

\bibitem{Llinares:2013qbh}
C.~Llinares and D.~Mota, ``{Releasing scalar fields: cosmological simulations
  of scalar-tensor theories for gravity beyond the static approximation},''
  \href{http://dx.doi.org/10.1103/PhysRevLett.110.161101}{{\em Phys.Rev.Lett.}
  {\bfseries 110} no.~16, (2013) 161101},
\href{http://arxiv.org/abs/1302.1774}{{\ttfamily arXiv:1302.1774
  [astro-ph.CO]}}.

\bibitem{Llinares:2013jua}
C.~Llinares and D.~F. Mota, ``{Cosmological simulations of screened modified
  gravity out of the static approximation: effects on matter distribution},''
  \href{http://dx.doi.org/10.1103/PhysRevD.89.084023}{{\em Phys.Rev.}
  {\bfseries D89} no.~8, (2014) 084023},
\href{http://arxiv.org/abs/1312.6016}{{\ttfamily arXiv:1312.6016
  [astro-ph.CO]}}.

\bibitem{Bose:2014zba}
S.~Bose, W.~A. Hellwing, and B.~Li, ``{Testing the quasi-static approximation
  in $f(R)$ gravity simulations},''
  \href{http://dx.doi.org/10.1088/1475-7516/2015/02/034}{{\em JCAP} {\bfseries
  1502} no.~02, (2015) 034},
\href{http://arxiv.org/abs/1411.6128}{{\ttfamily arXiv:1411.6128
  [astro-ph.CO]}}.

\bibitem{Amendola:2007rr}
L.~Amendola, M.~Kunz, and D.~Sapone, ``{Measuring the dark side (with weak
  lensing)},'' \href{http://dx.doi.org/10.1088/1475-7516/2008/04/013}{{\em
  JCAP} {\bfseries 0804} (2008) 013},
\href{http://arxiv.org/abs/0704.2421}{{\ttfamily arXiv:0704.2421 [astro-ph]}}.

\bibitem{Hu:2007pj}
W.~Hu and I.~Sawicki, ``{A Parameterized Post-Friedmann Framework for Modified
  Gravity},'' \href{http://dx.doi.org/10.1103/PhysRevD.76.104043}{{\em
  Phys.Rev.} {\bfseries D76} (2007) 104043},
\href{http://arxiv.org/abs/0708.1190}{{\ttfamily arXiv:0708.1190 [astro-ph]}}.

\bibitem{Bertschinger:2008zb}
E.~Bertschinger and P.~Zukin, ``{Distinguishing Modified Gravity from Dark
  Energy},'' \href{http://dx.doi.org/10.1103/PhysRevD.78.024015}{{\em
  Phys.Rev.} {\bfseries D78} (2008) 024015},
\href{http://arxiv.org/abs/0801.2431}{{\ttfamily arXiv:0801.2431 [astro-ph]}}.

\bibitem{Zhao:2008bn}
G.-B. Zhao, L.~Pogosian, A.~Silvestri, and J.~Zylberberg, ``{Searching for
  modified growth patterns with tomographic surveys},''
  \href{http://dx.doi.org/10.1103/PhysRevD.79.083513}{{\em Phys.Rev.}
  {\bfseries D79} (2009) 083513},
\href{http://arxiv.org/abs/0809.3791}{{\ttfamily arXiv:0809.3791 [astro-ph]}}.

\bibitem{Pogosian:2010tj}
L.~Pogosian, A.~Silvestri, K.~Koyama, and G.-B. Zhao, ``{How to optimally
  parametrize deviations from General Relativity in the evolution of
  cosmological perturbations?},''
  \href{http://dx.doi.org/10.1103/PhysRevD.81.104023}{{\em Phys.Rev.}
  {\bfseries D81} (2010) 104023},
\href{http://arxiv.org/abs/1002.2382}{{\ttfamily arXiv:1002.2382
  [astro-ph.CO]}}.

\bibitem{Bean:2010zq}
R.~Bean and M.~Tangmatitham, ``{Current constraints on the cosmic growth
  history},'' \href{http://dx.doi.org/10.1103/PhysRevD.81.083534}{{\em
  Phys.Rev.} {\bfseries D81} (2010) 083534},
\href{http://arxiv.org/abs/1002.4197}{{\ttfamily arXiv:1002.4197
  [astro-ph.CO]}}.

\bibitem{Sawicki:2012re}
I.~Sawicki, I.~D. Saltas, L.~Amendola, and M.~Kunz, ``{Consistent perturbations
  in an imperfect fluid},''
  \href{http://dx.doi.org/10.1088/1475-7516/2013/01/004}{{\em JCAP} {\bfseries
  1301} (2013) 004},
\href{http://arxiv.org/abs/1208.4855}{{\ttfamily arXiv:1208.4855
  [astro-ph.CO]}}.

\bibitem{Baker:2012zs}
T.~Baker, P.~G. Ferreira, and C.~Skordis, ``{The Parameterized Post-Friedmann
  Framework for Theories of Modified Gravity: Concepts, Formalism and
  Examples},'' \href{http://dx.doi.org/10.1103/PhysRevD.87.024015}{{\em
  Phys.Rev.} {\bfseries D87} (2013) 024015},
\href{http://arxiv.org/abs/1209.2117}{{\ttfamily arXiv:1209.2117
  [astro-ph.CO]}}.

\bibitem{Amendola:2012ky}
L.~Amendola, M.~Kunz, M.~Motta, I.~D. Saltas, and I.~Sawicki, ``{Observables
  and unobservables in dark energy cosmologies},''
  \href{http://dx.doi.org/10.1103/PhysRevD.87.023501}{{\em Phys.Rev.}
  {\bfseries D87} (2013) 023501},
\href{http://arxiv.org/abs/1210.0439}{{\ttfamily arXiv:1210.0439
  [astro-ph.CO]}}.

\bibitem{Silvestri:2013ne}
A.~Silvestri, L.~Pogosian, and R.~V. Buniy, ``{A practical approach to
  cosmological perturbations in modified gravity},''
  \href{http://dx.doi.org/10.1103/PhysRevD.87.104015}{{\em Phys.Rev.}
  {\bfseries D87} (2013) 104015},
\href{http://arxiv.org/abs/1302.1193}{{\ttfamily arXiv:1302.1193
  [astro-ph.CO]}}.

\bibitem{Motta:2013cwa}
M.~Motta, I.~Sawicki, I.~D. Saltas, L.~Amendola, and M.~Kunz, ``{Probing Dark
  Energy through Scale Dependence},''
  \href{http://dx.doi.org/10.1103/PhysRevD.88.124035}{{\em Phys.Rev.}
  {\bfseries D88} (2013) 124035},
\href{http://arxiv.org/abs/1305.0008}{{\ttfamily arXiv:1305.0008
  [astro-ph.CO]}}.

\bibitem{Creminelli:2008wc}
P.~Creminelli, G.~D'Amico, J.~Norena, and F.~Vernizzi, ``{The Effective Theory
  of Quintessence: the $w<-1$ Side Unveiled},''
  \href{http://dx.doi.org/10.1088/1475-7516/2009/02/018}{{\em JCAP} {\bfseries
  0902} (2009) 018},
\href{http://arxiv.org/abs/0811.0827}{{\ttfamily arXiv:0811.0827 [astro-ph]}}.

\bibitem{Gubitosi:2012hu}
G.~Gubitosi, F.~Piazza, and F.~Vernizzi, ``{The Effective Field Theory of Dark
  Energy},'' \href{http://dx.doi.org/10.1088/1475-7516/2013/02/032}{{\em JCAP}
  {\bfseries 1302} (2013) 032},
\href{http://arxiv.org/abs/1210.0201}{{\ttfamily arXiv:1210.0201 [hep-th]}}.

\bibitem{Bloomfield:2012ff}
J.~K. Bloomfield, E.~E. Flanagan, M.~Park, and S.~Watson, ``{Dark energy or
  modified gravity? An effective field theory approach},''
  \href{http://dx.doi.org/10.1088/1475-7516/2013/08/010}{{\em JCAP} {\bfseries
  1308} (2013) 010},
\href{http://arxiv.org/abs/1211.7054}{{\ttfamily arXiv:1211.7054
  [astro-ph.CO]}}.

\bibitem{Piazza:2013pua}
F.~Piazza, H.~Steigerwald, and C.~Marinoni, ``{Phenomenology of dark energy:
  exploring the space of theories with future redshift surveys},''
\href{http://arxiv.org/abs/1312.6111}{{\ttfamily arXiv:1312.6111
  [astro-ph.CO]}}.

\bibitem{Gleyzes:2013ooa}
J.~Gleyzes, D.~Langlois, F.~Piazza, and F.~Vernizzi, ``{Essential Building
  Blocks of Dark Energy},''
  \href{http://dx.doi.org/10.1088/1475-7516/2013/08/025}{{\em JCAP} {\bfseries
  1308} (2013) 025},
\href{http://arxiv.org/abs/1304.4840}{{\ttfamily arXiv:1304.4840 [hep-th]}}.

\bibitem{Bellini:2014fua}
E.~Bellini and I.~Sawicki, ``{Maximal freedom at minimum cost: linear
  large-scale structure in general modifications of gravity},''
  \href{http://dx.doi.org/10.1088/1475-7516/2014/07/050}{{\em JCAP} {\bfseries
  1407} (2014) 050},
\href{http://arxiv.org/abs/1404.3713}{{\ttfamily arXiv:1404.3713
  [astro-ph.CO]}}.

\bibitem{Gleyzes:2014dya}
J.~Gleyzes, D.~Langlois, F.~Piazza, and F.~Vernizzi, ``{Healthy theories beyond
  Horndeski},''
\href{http://arxiv.org/abs/1404.6495}{{\ttfamily arXiv:1404.6495 [hep-th]}}.

\bibitem{Horndeski:1974wa}
G.~W. Horndeski, ``{Second-order scalar-tensor field equations in a
  four-dimensional space},''
\href{http://dx.doi.org/10.1007/BF01807638}{{\em Int.J.Theor.Phys.} {\bfseries
  10} (1974) 363--384}.

\bibitem{Deffayet:2011gz}
C.~Deffayet, X.~Gao, D.~Steer, and G.~Zahariade, ``{From k-essence to
  generalised Galileons},''
  \href{http://dx.doi.org/10.1103/PhysRevD.84.064039}{{\em Phys.Rev.}
  {\bfseries D84} (2011) 064039},
\href{http://arxiv.org/abs/1103.3260}{{\ttfamily arXiv:1103.3260 [hep-th]}}.

\bibitem{Zumalacarregui:2013pma}
M.~Zumalacárregui and J.~García-Bellido, ``{Transforming gravity: from
  derivative couplings to matter to second-order scalar-tensor theories beyond
  the Horndeski Lagrangian},''
  \href{http://dx.doi.org/10.1103/PhysRevD.89.064046}{{\em Phys. Rev.}
  {\bfseries D89} (2014) 064046},
\href{http://arxiv.org/abs/1308.4685}{{\ttfamily arXiv:1308.4685 [gr-qc]}}.

\bibitem{Gao:2014soa}
X.~Gao, ``{Unifying framework for scalar-tensor theories of gravity},''
  \href{http://dx.doi.org/10.1103/PhysRevD.90.081501}{{\em Phys. Rev.}
  {\bfseries D90} (2014) 081501},
\href{http://arxiv.org/abs/1406.0822}{{\ttfamily arXiv:1406.0822 [gr-qc]}}.

\bibitem{Planck:2015xua}
{\bfseries Planck Collaboration} Collaboration, P.~Ade {\em et~al.}, ``{Planck
  2015 results. XIII. Cosmological parameters},''
\href{http://arxiv.org/abs/1502.01589}{{\ttfamily arXiv:1502.01589
  [astro-ph.CO]}}.

\bibitem{Gleyzes:2014rba}
J.~Gleyzes, D.~Langlois, and F.~Vernizzi, ``{A unifying description of dark
  energy},'' \href{http://dx.doi.org/10.1142/S021827181443010X}{{\em
  Int.J.Mod.Phys.} {\bfseries D23} (2014) 3010},
\href{http://arxiv.org/abs/1411.3712}{{\ttfamily arXiv:1411.3712 [hep-th]}}.

\bibitem{Blas:2015xga}
D.~Blas, ``{Lorentz violation in gravity},'' in {\em {50th Rencontres de
  Moriond on Gravitation: 100 years after GR La Thuile, Italy, March 21-28,
  2015}}.
\newblock 2015.
\newblock
\href{http://arxiv.org/abs/1507.07687}{{\ttfamily arXiv:1507.07687 [gr-qc]}}.
\newblock

\bibitem{Saltas:2014dha}
I.~D. Saltas, I.~Sawicki, L.~Amendola, and M.~Kunz, ``{Anisotropic Stress as a
  Signature of Nonstandard Propagation of Gravitational Waves},''
  \href{http://dx.doi.org/10.1103/PhysRevLett.113.191101}{{\em Phys.Rev.Lett.}
  {\bfseries 113} no.~19, (2014) 191101},
\href{http://arxiv.org/abs/1406.7139}{{\ttfamily arXiv:1406.7139
  [astro-ph.CO]}}.

\bibitem{Baker:2014zva}
T.~Baker, P.~G. Ferreira, C.~D. Leonard, and M.~Motta, ``{New Gravitational
  Scales in Cosmological Surveys},''
  \href{http://dx.doi.org/10.1103/PhysRevD.90.124030}{{\em Phys.Rev.}
  {\bfseries D90} no.~12, (2014) 124030},
\href{http://arxiv.org/abs/1409.8284}{{\ttfamily arXiv:1409.8284
  [astro-ph.CO]}}.

\bibitem{Tsujikawa:2007xu}
S.~Tsujikawa, ``{Observational signatures of f(R) dark energy models that
  satisfy cosmological and local gravity constraints},''
  \href{http://dx.doi.org/10.1103/PhysRevD.77.023507}{{\em Phys.Rev.}
  {\bfseries D77} (2008) 023507},
\href{http://arxiv.org/abs/0709.1391}{{\ttfamily arXiv:0709.1391 [astro-ph]}}.

\bibitem{Kunz:2015oqa}
M.~Kunz, S.~Nesseris, and I.~Sawicki, ``{Using dark energy to suppress power at
  small scales},'' \href{http://dx.doi.org/10.1103/PhysRevD.92.063006}{{\em
  Phys. Rev.} {\bfseries D92} no.~6, (2015) 063006},
\href{http://arxiv.org/abs/1507.01486}{{\ttfamily arXiv:1507.01486
  [astro-ph.CO]}}.

\bibitem{Samushia:2013yga}
L.~Samushia, B.~A. Reid, M.~White, W.~J. Percival, A.~J. Cuesta, {\em et~al.},
  ``{The Clustering of Galaxies in the SDSS-III Baryon Oscillation
  Spectroscopic Survey (BOSS): measuring growth rate and geometry with
  anisotropic clustering},'' \href{http://dx.doi.org/10.1093/mnras/stu197}{{\em
  Mon.Not.Roy.Astron.Soc.} {\bfseries 439} (2014) 3504--3519},
\href{http://arxiv.org/abs/1312.4899}{{\ttfamily arXiv:1312.4899
  [astro-ph.CO]}}.

\bibitem{Doran:2006kp}
M.~Doran and G.~Robbers, ``{Early dark energy cosmologies},''
  \href{http://dx.doi.org/10.1088/1475-7516/2006/06/026}{{\em JCAP} {\bfseries
  0606} (2006) 026},
\href{http://arxiv.org/abs/astro-ph/0601544}{{\ttfamily arXiv:astro-ph/0601544
  [astro-ph]}}.

\bibitem{Pettorino:2013ia}
V.~Pettorino, L.~Amendola, and C.~Wetterich, ``{How early is early dark
  energy?},'' \href{http://dx.doi.org/10.1103/PhysRevD.87.083009}{{\em
  Phys.Rev.} {\bfseries D87} (2013) 083009},
\href{http://arxiv.org/abs/1301.5279}{{\ttfamily arXiv:1301.5279
  [astro-ph.CO]}}.

\bibitem{Song:2006jk}
Y.-S. Song, I.~Sawicki, and W.~Hu, ``{Large-Scale Tests of the DGP Model},''
  \href{http://dx.doi.org/10.1103/PhysRevD.75.064003}{{\em Phys.Rev.}
  {\bfseries D75} (2007) 064003},
\href{http://arxiv.org/abs/astro-ph/0606286}{{\ttfamily arXiv:astro-ph/0606286
  [astro-ph]}}.

\bibitem{Ade:2015zua}
{\bfseries Planck} Collaboration, P.~A.~R. Ade {\em et~al.}, ``{Planck 2015
  results. XV. Gravitational lensing},''
\href{http://arxiv.org/abs/1502.01591}{{\ttfamily arXiv:1502.01591
  [astro-ph.CO]}}.

\bibitem{Lewis:2006fu}
A.~Lewis and A.~Challinor, ``{Weak gravitational lensing of the cmb},''
  \href{http://dx.doi.org/10.1016/j.physrep.2006.03.002}{{\em Phys. Rept.}
  {\bfseries 429} (2006) 1--65},
\href{http://arxiv.org/abs/astro-ph/0601594}{{\ttfamily arXiv:astro-ph/0601594
  [astro-ph]}}.

\bibitem{Simpson:2012ra}
F.~Simpson, C.~Heymans, D.~Parkinson, C.~Blake, M.~Kilbinger, {\em et~al.},
  ``{CFHTLenS: Testing the Laws of Gravity with Tomographic Weak Lensing and
  Redshift Space Distortions},''
  \href{http://dx.doi.org/10.1093/mnras/sts493}{{\em Mon.Not.Roy.Astron.Soc.}
  {\bfseries 429} (2013) 2249},
\href{http://arxiv.org/abs/1212.3339}{{\ttfamily arXiv:1212.3339
  [astro-ph.CO]}}.

\bibitem{Abbott:2005bi}
{\bfseries Dark Energy Survey Collaboration} Collaboration, T.~Abbott {\em
  et~al.}, ``{The dark energy survey},''
\href{http://arxiv.org/abs/astro-ph/0510346}{{\ttfamily arXiv:astro-ph/0510346
  [astro-ph]}}.

\bibitem{Amendola:2013qna}
L.~Amendola, S.~Fogli, A.~Guarnizo, M.~Kunz, and A.~Vollmer,
  ``{Model-independent constraints on the cosmological anisotropic stress},''
  \href{http://dx.doi.org/10.1103/PhysRevD.89.063538}{{\em Phys. Rev.}
  {\bfseries D89} no.~6, (2014) 063538},
\href{http://arxiv.org/abs/1311.4765}{{\ttfamily arXiv:1311.4765
  [astro-ph.CO]}}.

\bibitem{Hu:2013twa}
B.~Hu, M.~Raveri, N.~Frusciante, and A.~Silvestri, ``{Effective Field Theory of
  Cosmic Acceleration: an implementation in CAMB},''
  \href{http://dx.doi.org/10.1103/PhysRevD.89.103530}{{\em Phys. Rev.}
  {\bfseries D89} no.~10, (2014) 103530},
\href{http://arxiv.org/abs/1312.5742}{{\ttfamily arXiv:1312.5742
  [astro-ph.CO]}}.

\end{thebibliography}\endgroup

\end{document}